%% file: IJCNN23.tex
\documentclass[conference]{IEEEtran}
\IEEEoverridecommandlockouts

\usepackage{cite}
\usepackage{amsmath,amssymb,amsfonts}
\usepackage{algorithmic}
\usepackage{graphicx}
\usepackage{textcomp}
\usepackage{xcolor}
\def\BibTeX{{\rm B\kern-.05em{\sc i\kern-.025em b}\kern-.08em
    T\kern-.1667em\lower.7ex\hbox{E}\kern-.125emX}}

\usepackage{amsthm, xspace, color, multirow, mathrsfs, booktabs}
\usepackage{algorithm}

\newcommand{\etc}{\emph{etc.}\xspace} 

\newcommand{\nop}[1]{}

\begin{document}

\title{Market Making with Deep Reinforcement Learning from Limit Order Books}

\author{
\IEEEauthorblockN{Hong Guo}
\IEEEauthorblockA{\textit{Shenzhen International Graduate School} \\
\textit{Tsinghua University}\\
Shenzhen, China \\
guoh20@mails.tsinghua.edu.cn}
\and
\IEEEauthorblockN{Jianwu Lin \thanks{\IEEEauthorrefmark{1} Corresponding author.}\IEEEauthorrefmark{1}}
\IEEEauthorblockA{\textit{Shenzhen International Graduate School} \\
\textit{Tsinghua University}\\
Shenzhen, China \\
Lin.Jianwu@sz.tsinghua.edu.cn}
\and
\IEEEauthorblockN{Fanlin Huang}
\IEEEauthorblockA{
\textit{Microsoft} \\
Beijing, China \\
fanlinghuang@microsoft.com}
}

\maketitle

\input{sections/abstract.tex}


\input{sections/intro}
\input{sections/preliminaries}

\input{sections/method}
\input{sections/experiment}

\input{sections/conclusion}

\bibliographystyle{IEEEtran}
\bibliography{sample-base}

\end{document}

%% file: sections/abstract.tex
\begin{abstract}
  Market making (MM) is an important research topic in quantitative finance, the agent needs to continuously optimize ask and bid quotes to provide liquidity and make profits.
  The limit order book (LOB) contains information on all active limit orders, which is an essential basis for decision-making.
  The modeling of evolving, high-dimensional and low signal-to-noise ratio LOB data is a critical challenge.  
  Traditional MM strategy relied on strong assumptions such as price process, order arrival process \etc
  Previous reinforcement learning (RL) works handcrafted market features, which is insufficient to represent the market.
  This paper proposes a RL agent for market making with LOB data. We leverage a neural network with convolutional filters and attention mechanism (Attn-LOB) for feature extraction from LOB. We design a new continuous action space and a hybrid reward function for the MM task. Finally, we conduct comprehensive experiments on latency and interpretability, showing that our agent has good applicability.
\end{abstract}

%% file: sections/intro.tex
\section{Introduction}

Market making (MM) is a salient problem in the quantitative finance domain. Market makers usually quote limit orders on both ask and bid sides to capture spreads and make profits. Market makers, who provide liquidity and immediacy to the market, play a crucial role in the price discovery process. 

While market makers are faced with three main costs: adverse information costs (45\%), order processing costs (45\%), and inventory holding costs (10\%) \cite{stoll1989inferring}. 
Adverse information costs are the cost paid by uninformed traders to informed traders. 
Order processing costs refer to commission fees, such as stamp duty, transfer fees, \etc
Inventory holding costs are due to inventory, which could suffer losses when the value of the asset fluctuates.


More than half of the markets in today’s financial world use an electronic limit order book (LOB) to facilitate trade \cite{rocsu2009dynamic}. 
 The LOB is a set of all active orders submitted to the Exchange. Cao et al. \cite{cao2009information} found that the LOB is informative and contributes approximately 22\% to price discovery.
 
 The challenge is that the state space of a LOB is huge, 
which makes it very difficult to investigate conditional dependencies.
Therefore, a key modeling task is to find a way to simplify the evolving, high-dimensional state space, while retaining LOB’s important features \cite{gould2013limit}. 
In an electronic matching market, to develop an autonomous market making agent has become a practical problem. 

This paper presents an adaptive agent for market making under a simulated environment. \footnote{The source code is available at https://github.com/imTurkey/Market-Making-with-Deep-Reinforcement-Learning-from-Limit-Order-Books}
A CNN-Attention based network (Attn-LOB) is proposed as a function approximator to extract features from LOB.
There are three main challenges, the first is the feature extraction from the noisy LOB data, the second is to define a suitable action space for market making, and the last is to reward the agent properly to the right decision, which is a challenging task even for a human trader.

\subsection{Related Work}
For market making, Demsetz \cite{demsetz1968cost} first proposed a theory, he pointed out the bid-ask spread is the compensation that markets provide for the immediacy of transactions.
Then, Garman et al. \cite{garman1976market} proposed an inventory based model for market making. Afterwards, the model developed cases from a single market maker to multiple market makers \cite{stoll1978pricing,ho1981optimal}. 
Many theoretical market making models are developed in the context of stochastic dynamic programming. 
A widely used model was proposed by Avellaneda and Stoikov \cite{avellaneda2008high}, who calibrated the optimal quotes by considering the probability with which his quotes will be executed as a function of their distance from the mid-price, this work followed Ho and Stoll \cite{ho1980dealer,ho1981optimal}. Avellaneda-Stoikov (AS) model assumes that the price is continuous, which is not true because of the minimum price tick. Guilbaud et al. \cite{guilbaud2013optimal} constructed a strategy for discrete quotes.
The main limitation of these models is that specific properties of the underlying processes (price process and order arrival process) have to be assumed in order to obtain a closed-form characterization of strategies \cite{chan2001electronic}.

The ever rising speed and decreasing costs of computational power and networks have led to the emergence of huge databases that record all transactions and order book movements up to milliseconds \cite{chakraborti2011econophysics}. Many econophysics literature \cite{garman1976market,kyle1985continuous,glosten1994electronic,o1998market} studied the microstructure of LOBs. 
In recent years, with the popularity of data-driven methods, a lot of data-driven methods \cite{sirignano2019universal,tsantekidis2020using,ntakaris2019feature,tran2017tensor} tried to extract features from noisy financial data.
And machine learning methods \cite{nousi2019machine,zhang2019extending,zhang2019deeplob,zhang2021multi,ntakaris2018benchmark} were applied to short-term price forecasting. 
Also, deep learning methods such as Convolutional Neural Network (CNN) \cite{tsantekidis2017forecasting}, Recurrent Neural Network (RNN) \cite{tsantekidis2017using} and Attention \cite{tran2018temporal,shabani2022multi} were leveraged to predict price movement and discover patterns in the high-frequency trading (HFT) domain.

The above methods were mainly used to predict signals. Many RL methods were used for trading \cite{moody1998reinforcement,moody2001learning,deng2016deep,wei2019model,briola2021deep}, optimal execution \cite{nevmyvaka2006reinforcement} and portfolio management \cite{wang2021deeptrader,wang2021commission,jin2016portfolio}. 

Many work applied RL to market making,
Abernethy \cite{abernethy2013adaptive} is the first to leverage online learning to solve market making problem.
Then, Lim and Gorse \cite{lim2018reinforcement} and Spooner et al. \cite{spooner2018market} applied non-deep RL for simulated and real-world historic data respectively, \cite{lim2018reinforcement} used inventory and remaining time to describe the state, \cite{spooner2018market} took handcrafted LOB features such as spread, price movement, order imbalance into account, their action spaces are both discrete.
Zhong \cite{zhong2020data} handcrafted features to characterize market volatility and mid-price movement, but their action space allowed the agent to exit the market.
Sadighian \cite{sadighian2019deep} first leveraged Deep Reinforcement Learning (DRL) in cryptocurrency market making, he used LOB, TFI (Trade Flow Imbalance) and OFI (Order Flow Imbalance) to describe the market state, he leveraged a MLP as the function approximator, which could have limited the ability of LOB description. 
Gašperov and Kostanjčar \cite{gavsperov2021market} used a gradient boosting model to predict realized price range and a long short-term memory (LSTM) to predict the trend, and the signals were used to train a DRL agent.
Some latest researches \cite{kumar2020deep, xu2022performance} leveraged Deep Recurrent Q-Networks (DRQN) and Dueling Double Deep Q Network (D3QN) respectively, however, neither of them considered the characteristics of LOB while designing deep networks.


These work have either used simulated market data, handcrafted features  
or used a simple MLP for the feature extraction from LOBs,
which could have restricted the market-making ability of reinforcement learning agents.


\subsection{Contributions}

The main contribution of this paper is the design of an adaptive agent for market making with reinforcement learning from limit order book. In the financial domain, many scenario-specific metrics such as latency, inventory and Sharpe are investigated, and a simulated backtesting paradigm of RL for market making is established. In contrast to previous MM models \cite{avellaneda2008high,guilbaud2013optimal} which rely on strong assumptions, we use a model-free approach which need not to explicitly model the environment.
Compared with the previous RL agents \cite{spooner2018market,mani2019applications,lim2018reinforcement,gavsperov2021market,abernethy2013adaptive} which handcrafted market representation, we extract abundant stationary and dynamic market features from LOB data, which is proved effective for market making. 
Our main contributions can be summarized below:

\begin{itemize}
    \item We propose an effective framework to train an RL agent with automatic feature extraction and pre-training from LOB data for market making problems, which is a novel challenge in the HFT domain.
    \item We propose a CNN-Attention based network (Attn-LOB) for automatic feature extraction from LOB data, we pre-trained it in a mid-price direction prediction task, Attn-LOB is proved effective for establishing the representation of the market. 
    \item We propose a novel continuous action space close to the financial practice for training the RL agent, we compared it to a previous discrete action space which limits the agent to make decisions. 
    \item We study several reward functions designed for capturing spreads and controlling inventory in theory and experiments. We design a hybrid reward function for market making and 
    show their effectiveness. 
    \item We conduct an interpretability experiment to explain our agent, we find that the agent can pay attention to the relevant market changes and learn the basic skills of market making, which showed great practical potential.

\end{itemize}

%% file: sections/preliminaries.tex
\section{Preliminaries}

\subsection{Limit Order Book}


There are two types of orders: Market Orders and Limit Orders. 
A market order is an order to buy or sell at the market's current best available price, which typically ensures an execution but does not guarantee a specified price. It is executed against limit orders starting with the best price. In a less-liquid market, a market order may spill over to further price levels if there is not enough volume at the current level of the order book. 
A limit order is an order to buy or sell with a restriction on the maximum price to be paid or the minimum price to be received (the "limit price"), it does not assure execution. 
When a limit order arrives at the Exchange, it joins the back of the order queue at its quoted price level and will be filled on a first-come first-served rule. 
The collection of all active limit orders is a limit order book (LOB). 

The state space of a limit order book is inherently high-dimensional, posing a significant challenge to modeling efforts. 
For instance, if there exist $P$ price levels, the simplest aggregated state space would be $\mathbb{Z}^{P}$, which renders the modeling process complex. 
Furthermore, the limit order book is often characterized by noise, owing to the presence of numerous irrelevant signals, such as quoting and immediately canceling orders.

\subsection{Market Making}

A market maker (MMer) is a company or an individual that quotes both buy and sell limit orders in an asset to make a profit. An MMer's goal is to earn the difference between the quoted bid and ask prices. If both orders get executed, then the MMer will gain the quoted spread. 
In an ideal case, the inventory-based MMer seeks to get both orders executed to minimize the inventory risk. But there is a possible scenario that only one of these orders gets executed, which results in the MMer getting a non-zero inventory, thereby exposing itself to the inventory risk. When faced with inventory risk, the MMer typically skews its quotes in order to reduce its inventory position. 

\subsection{Reinforcement Learning}

Reinforcement learning (RL) is a machine learning method used to solve sequential decision problems. In the trading domain, RL provides a way to directly output decisions using an end-to-end training process rather than a two-stage prediction and decision-making process. Since the decision targets are difficult to label, the supervised learning methods are difficult to implement, and RL approaches provide a new perspective to train agents through reward signals.

%% file: sections/method.tex
\section{Method}

\subsection{Market Representation}

\subsubsection{Order Book State}

The snapshot of an order book at time \textit{t} can be simplified to a one-dimensional vector whose length is $4*n$, where $n$ represents the number of levels.
\begin{equation}
    LOB_t = \left\{P_{i,t}^{\text {ask }}, V_{i,t}^{\text {ask }}, P_{i,t}^{\text {bid }}, V_{i,t}^{\text {bid }}\right\}_{i=1}^n
\end{equation}

Where $P_{i,t}$ denotes the price of level $i$ at time $t$, $V_{i,t}$ denotes the total volume of level $i$ at time $t$ . 
We adopt a rolling window of length $T$ to form the input, the shape of which is ($T$, 4*$n$).
Handcrafted features are limited and insufficient to describe a high-frequency market. 
We pre-trained a deep neural network and applied this model as a function approximator 
during reinforcement learning, which will be explained in Section \ref{sec:auto_feature_extract}. 

\subsubsection{Dynamic market state}

The limit order book only contains stationary information. In this section, we design a new indicator and leverage two commonly used indicators to obtain the dynamic market states.

\begin{itemize}

    \item \textbf{Order Strength Index} (OSI). To consider the dynamic impact of newly happened events on the market, we customize OSI to describe the relative strength between the bid and ask. The events can be divided into three categories: new market orders, new limit orders and order cancellations. The strength indexes in both volumes and numbers of each category are calculated by:
    \begin{equation}
        OSI_{v}=\frac{\sum V_{buy}-\sum V_{sell}}{\sum V_{buy}+\sum V_{sell}} 
    \end{equation}

    where $V$ denotes the volume, and we also calculate OSI based on the number of orders.
    OSI is calculated on the time window of 10s, 60s and 300s respectively to obtain 18 features.
    
    \item \textbf{Realized Volatility} (RV) is an assessment of variation for assets by analyzing its historical returns within a defined period, it can be calculated by:
    \begin{equation}
        RV = \sqrt{\sum_{t=1}^{n}{(\log_{}{p_t} - \log_{}{p_{t-1}})}^2 } 
    \end{equation}
    where $p_{t}$ denotes the price at time $t$. We calculate the RV for the past 5 minutes, 10 minutes, and 30 minutes respectively, and get 3 features.
    
    \item \textbf{Relative Strength Index} (RSI) \cite{wilder1978new} is a technical indicator used in momentum trading that measures the speed of a security's recent price changes to evaluate overvalued or undervalued conditions in the price of that security. It can be calculated by:
    \begin{equation}
        RSI = \frac{\sum_{t=1}^{n} Gain_t}{\sum_{t=1}^{n} Gain_t+\sum_{t=1}^{n} Loss_t} 
    \end{equation}
    

    Where $Gain_t = max(0,p_t - p_{t-1})$, $Loss_t = max(0,p_{t-1}-p_t)$. Similarly, we calculate the RSI in 5 minutes, 10 minutes, and 30 minutes respectively, and obtain 3 features.

\end{itemize} 

\subsubsection{Agent State}

Agent state contains inventory and time factor. \textbf{Inventory} is a critical internal state to control the inventory risk.
\textbf{Remaining Time} is calculated by current time $t$ divided by total time $T$.


\subsection{Limit Order Book Modeling}
\label{sec:auto_feature_extract}

We use a CNN-Attention based network to extract features from the LOB data, and we use pre-train \cite{mikolov2013efficient} to improve its performance. The mid-price direction prediction is a widely studied problem and can be used as the pre-train task.
The task is to classify the future price trend (up, down or stationary) in a certain horizon with limit order book of the past $t$ timestamps.

\subsubsection{Label Acquisition}
\label{seq:label}

We use the approach proposed in \cite{zhang2019deeplob} to obtain labels, which are proved to be more robust:
\begin{equation}
y_t=\begin{cases}1& \text{if}\ l_t > \alpha
 \\0& \text{if}\ -\alpha \le  l_t \le  \alpha
 \\-1& \text{if}\ l_t < -\alpha
\end{cases}
\end{equation}
where $\alpha$ is a threshold hyperparameter which can be set according to the market. And $l_t$ can be calculated by
\begin{equation}
l_t=\frac{m_+(t) - m_-(t)}{m_-(t)} 
\end{equation}
\begin{equation}
m_\pm(t)=\frac{1}{k}\sum_{i=0}^{k}p_{t\mp i}
\end{equation}

Where $p_t$ is the mid-price and $k$ denotes the predicted horizon, 

\subsubsection{Normalization}
Since the value of an asset may fluctuate to a never before seen level, statistics of the price values can change significantly with time, rendering the price time series non-stationary.
In order to transform the non-stationary input sequence into a stationary sequence, we adopt normalization described in \cite{tsantekidis2020using}.
Afterwards, we perform \textit{z-norm} to the stationary price sequence and \textit{max-norm} for the volume sequence.

\subsubsection{Model}
\label{sec:model}

\begin{figure*}[h]
\setlength{\abovecaptionskip}{-0.1cm}
 \setlength{\belowcaptionskip}{-0.3cm}
  \centering
  \includegraphics[width=18cm]{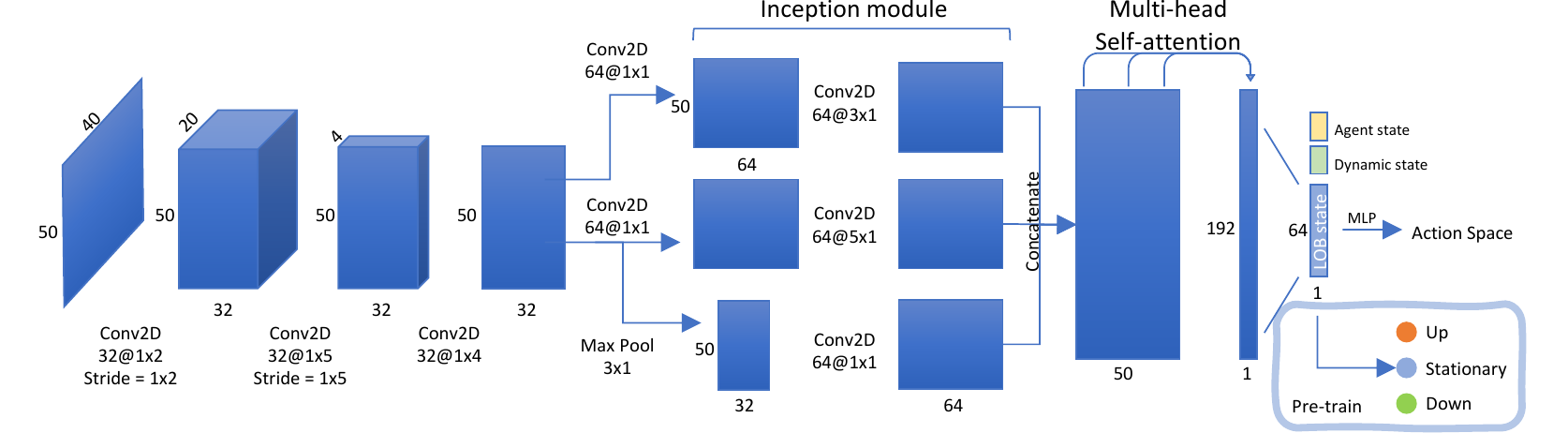}
  \caption{The model architecture.}
  \label{fig:network}
\end{figure*}

The model we use, Attn-LOB, is a deep neural network which is composed by convolutional networks and self-attention. 
The model architecture follows \cite{ntakaris2019feature, zhang2019deeplob}.
They proposed a CNN-LSTM based network and
 we add a \textit{multi-head self-attention} layer to model temporal dependencies. The network architecture is shown in Figure \ref{fig:network}.
The input shape is $T\times(4*n)$. 
First, multiple convolutional filters are used to model the spatial dependencies. Then the output in shape $T\times hidde\_dim$ are fed into an \textit{Inception} module to establish multi-scale temporal dependencies. At last, a \textit{multi-head self-attention} layer was used to aggregate  temporal dependencies.

\subsection{Action Space}
A typical market making strategy is to submit limit orders on both buy and sell sides. The agent is restricted to a single buy and sell order and cannot exit the market, which is similar to \cite{chakraborty2011market}. Here we introduce two action spaces. The first is discrete and has been used many times in past work. The second is a continuous action space designed by us.

\subsubsection{Discrete}


The discrete action space $A$ is similar to \cite{spooner2018market,sadighian2019deep} and consists of 8 possible actions.
Action 0-7 is to quote a pair of orders with a particular spread and bias, and action 7 is to close position with market orders. 
The ask and bid prices will be calculated based on the price when orders are submitted. The volume of each order is the $minimum\_trade\_unit$, which is 100 in the China Stock Market.

\subsubsection{Continuous}


The continuous action space is more flexible than the discrete action space mentioned above. This type of action space is inspired by the AS strategy \cite{avellaneda2008high}, which obtains optimal ask and bid quotes by calculating a reservation price $p_r$ and a quotes spread, $p_r$ can be seen a "true" price of the asset \cite{ho1981optimal}, and the agent symmetrically quotes orders around $p_r$.

Action $A_1$ and $A_2$ are between 0 and 1. $A_1$ controls the bias $\delta$ of reservation price $p_r$ and mid-price $p_m$:
\begin{equation}
\label{eq:a1}
\delta = A_1*max\_bias
\end{equation}
Where $max\_bias$ is a hyper-parameter to decide the maximum bias of reservation price and mid-price. 
Then reservation price $p_r$ can be calculated by:
\begin{equation}
p_r = p_{m} - \text{Sign}(Inventory) * \delta
\end{equation}
The bias direction of reservation price is controlled by inventory, because we want the agent to buy when its inventory is negative and sell when its inventory is positive. 

Action $A_2$ controls the qouted spread:
\begin{equation}
spread = A_2*max\_spread
\end{equation}
where $max\_bias$ and $max\_spread$ are hyper-parameters to decide the maximum price bias and the maximum quotes spread respectively. The final quoted ask and bid prices are:
\begin{equation}
p_{a,b} = p_{r} \pm spread/2
\end{equation}


To avoid unlimited buying and selling, we set a hyper-parameter $\omega$ , when the agent's position is higher than $\omega * minimum\_trade\_unit$, the agent will be prohibited from placing orders in that direction.

\subsection{Reward Function}

\subsubsection{Dampened PnL}

The reward function is the key to guiding reinforcement learning. A natural thinking is to use Profit and Loss (PnL) as the reward function, which refers to the difference of value of the agent in $\Delta t$:
\begin{equation}
\Delta PnL_t=value_t -value_{t-\Delta t}
\end{equation}
Where value is calculated by the sum of inventory's value (calculated with current mid-price) and cash. Spooner et al. \cite{spooner2018market} has proved that only using PnL as the reward function will lead to the agent's speculative action. 
The agent tend to hold a large amount of inventory, which brings a great risk of drawdown when the trend market comes.
To avoid this, they added an asymmetric punishment term to calculate the Dampened PnL (DP):
\begin{equation}
DP_t = \Delta PnL_t - max(0,\eta * \Delta PnL_t)
\end{equation}
Where $\eta$ is a hyper-parameter. DP will reduce the reward for profit from holding, but not the punishment for loss.

\subsubsection{Trading PnL}

Trading PnL (TP) rewards the agent only at transaction $X_t=(X_{t,p},X_{t,v})$, where $X_{t,p}$ denotes the transaction price and $X_{t,v}$ denotes the transaction volume. TP is defined as:
\begin{equation}
TP_t=X_{t,v}*(p_{m,t} - X_{t,p})
\end{equation}
Where $p_{m,t}$ denotes the mid-price at time $t$. $X_{t,v}$ is positive for buying and negative for selling. It can be viewed as an advantage over the mid-price when trading. We can prove that $Trading\ PnL$ is equivalent to $\Delta PnL$ subtract $Holding\ PnL$, the latter is PnL caused by positions, defined by $Holding\ PnL = Inv_{t-\Delta t}*(p_{m,t}-p_{m,t-\Delta t})$. In short, $Trading\ PnL$ can reward the price advantage of trading rather than the profit or loss of inventory.




\subsubsection{Inventory Punishment} 

The inventory punishment (IP) is an effective way to control inventory risk:
\begin{equation}
\label{equa:inventory_punishment}
IP_t = \zeta * Inv_t^2 
\end{equation}

Where $\zeta$ is a hyper-parameter. In order to make the inventory punishment quickly increase, we use $L_2$ norm to calculate IP.

\subsubsection{Reward Function}
Finally, we combine these rewards above to formulate a hybrid reward function as Equation (\ref{equa:reward_function}). $DP_t$ are used to punish loss from holding inventory, $TP_t$ are used to reward advantageous price of trading, and $IP_t$ are used to punish the high inventory position.
\begin{equation}
\label{equa:reward_function}
R_t = DP_t + TP_t - IP_t
\end{equation}

\subsection{Training Algorithm}
We adopt two commonly used RL algorithms to train our agent: Dueling DQN is a value-based SOTA, and Proximal Policy Optimization (PPO) is a policy-based SOTA.


%% file: sections/experiment.tex
\section{Experiments}

\subsection{Dataset}
The data we use are historical orders and trades on the Shenzhen Stock Exchange in November 2019, covering 21 trading days. Data are updated event-by-event and an event may occur due to anything from a change in price, volume or arrangement of orders. We select three stocks in different industries to study the applicability of proposed agent, which are \textit{Ping An Bank Co.,Ltd.} (SZ.000001) in banking, \textit{Wuliangye Yibin Co.,Ltd.} (SZ.000858) in liquor-making, and \textit{Hikvision Technology Co.,Ltd.} (SZ.002415) in technology. We reconstruct their historical limit order books, which contain about 5,000,000 samples. 

\subsection{Pre-training}

\subsubsection{Experimental Setup}
\label{sec:exp_setup}

We conduct our experiments on Ubuntu 20.04 LTS. The machine consists of an Intel Xeon(R) E5-2650 v4 CPU and 48GB DRAM. The training data is the first half of the month (10 days), and the test data is the second half of the month (11 days), with 20\% of the training data used for validation.
We followed the setting of \cite{ntakaris2018benchmark} and only took data between 10:00 and 11:30, 13:00 and 14:30, during which trading was considered stable. We labeled these data as described in Section \ref{seq:label}.
We set the predicted horizon $k$ to 10 events, and the label threshold $\alpha$=1e-5, and the windows length $T=50$.

\subsubsection{Baselines}

FC-LOB is a multi-layer perception network. The number of neurons in the hidden layer is (1024, 256, 64, 3), and the activation function is leaky Relu except for Softmax in the last layer.
Conv-LOB is a fully convolutional network that uses dilated convolution to accept longer sequences. The architecture is similar to \cite{oord2016wavenet}.
DeepLOB is the network proposed by \cite{zhang2019deeplob}.
And Attn-LOB is our model described in Section \ref{sec:model}.

\subsubsection{Pre-train Results}

\begin{table}[]
\setlength{\abovecaptionskip}{-0.01cm}
 \setlength{\belowcaptionskip}{-0.01cm}
\caption{Pre-train results.}
\label{tab:pretrain}
\centering
\begin{tabular}{lccccc}
\toprule[1pt]
         & Precision       & Recall          & F1              & Param  & Input    \\ \hline
FC-LOB   & 0.6315          & 0.5419          & 0.5660          & 256,064 & $4000\times 1$  \\
Conv-LOB & 0.5851          & 0.5230          & 0.4984          & 172,320 & $1024\times 40$ \\
DeepLOB  & \textbf{0.7856} & 0.6699          & 0.7118          & 139,168 & $100\times 40$  \\
Attn-LOB & 0.7663          & \textbf{0.7019} & \textbf{0.7284} & 176,320 & $50\times 40$   \\ 
\bottomrule[1pt]
\end{tabular}
\end{table}

We present the results of pre-training on one stock ($Ping\ An\ Bank\ Co.,Ltd.$), and we also note that the model still performs well on stocks out of sample. As shown in Table \ref{tab:pretrain}, Attn-LOB outperforms other methods on recall and F1 score, and we only use half the time length as input compared with DeepLOB. Conv-LOB does not achieve the desired result, suggesting that we may not need to input such a long time sequence.

\subsection{RL Settings}
\subsubsection{Simulator}

We set up a simulated trading environment with the historical limit order book. To ensure the reality of simulated transactions, the simulator executes the agent's order only when the real historical order arrives. When the agent's bid price is higher than the lowest market ask price, or the agent's ask price is lower than the highest market bid price, the order gets executed and then the agent's cash and inventory will be updated. The transaction cost is set to 0, and the agent can hold negative cash or inventory. If the absolute inventory exceeds $max\_inventory=\omega * minimum\_trade\_unit$, the agent will be prohibited from quoting in that direction. If the agent chooses to close positions with market orders, the order will be executed at the counterparty's price. Since the volume of agent's quote is small, we ignore the impact of these orders on the market. 

    

\begin{table*}[]
\setlength{\abovecaptionskip}{-0.01cm}
 \setlength{\belowcaptionskip}{-0.01cm}
\caption{Overall results.}
\label{table:overall_results}
\centering
\scalebox{0.95}{
\begin{tabular}{lcccc|cccc|cccc}
\toprule[1pt]
\multicolumn{1}{c}{} & \multicolumn{4}{c}{Ping An Bank Co.,Ltd.}                                                                                 & \multicolumn{4}{c}{Wuliangye Yibin Co.,ltd.}                                                                              & \multicolumn{4}{c}{Hikvision Technology Co., Ltd.}                                                                        \\ \cline{2-13} 
\multicolumn{1}{c}{} & \begin{tabular}[c]{@{}c@{}}ND-PnL\\ $(\times 10^5)$\end{tabular} & PnLMAP & \begin{tabular}[c]{@{}c@{}}PR\\ $(\times 10^{-4})$\end{tabular} & Sharpe & \begin{tabular}[c]{@{}c@{}}ND-PnL\\ $(\times 10^5)$\end{tabular} & PnLMAP & \begin{tabular}[c]{@{}c@{}}PR\\ $(\times 10^{-4})$\end{tabular} & Sharpe & \begin{tabular}[c]{@{}c@{}}ND-PnL\\ $(\times 10^5)$\end{tabular} & PnLMAP & \begin{tabular}[c]{@{}c@{}}PR\\ $(\times 10^{-4})$\end{tabular} & Sharpe \\ \hline
C-PPO                  & 9.3±0.7                                                & 117.2±3.8  & 5.0±0.1                                             & 12.3±0.8   & 19.8±1.8                                                & 630.6±85.5 & 2.8±0.5                                             & 2.2±0.7   & 16.0±3.3                                                & 313.6±25.9  & 3.8±0.6                                             & 7.1±0.5   \\
D-DQN          & 7.0±1.7                                                 & 8.6±2.2  & 3.5±0.2                                             & 1.3±0.7   & 11.0±3.7                                                 & 28.4±12.8  & 0.9±0.1                                             & -0.5±0.1   & 11.7±6.8                                                 & 65.2±10.0  & 10.1±3.3                                             & 0.4±0.1   \\
Inv-RL          & 0.3±0.1                                                 & 24.7±4.2  & 4.3±0.9                                             & 1.3±0.3   & 3.8±0.4                                                 & 70.2±23.0  & 0.7±0.1                                             & -1.3±0.2   & 1.4±0.1                                                 & 52.7±16.6  & 2.4±0.4                                             & 1.0±0.4   \\
LOB-RL          & 1.1±0.5                                                 & 1.3±0.5  & 2.8±1.4                                             & 0.2±0.3   & 1.8±1.1                                                 & 6.8±5.0  & 0.2±0.1                                             & -0.7±0.3   & 1.9±0.9                                                 & 5.9±3.9  & 0.7±0.4                                             & -0.2±0.4   \\
AS                   & 0.49                                                 & 4.75   & 4.22                                             & 0.74   & 3.14                                                 & 19.61  & 3.93                                             & 0.17   & 1.57                                                 & 16.39  & 8.10                                             & 0.65   \\
Random               & 0.39                                                 & 0.81   & 0.93                                             & -0.19   & 0.86                                                 & 3.33   & 0.15                                             & -0.81   & 2.76                                                 & 6.73   & 1.75                                             & 0.27   \\
Fixed\_1                 & 2.63                                                 & 4.70   & 1.28                                             & -0.01   & -3.12                                                & -10.72 & -0.12                                            & -4.88  & 1.43                                                 & 3.71   & 0.24                                             & -1.69   \\
Fixed\_2                 & 0.97                                                 & 2.03   & 9.97                                             & 0.21   & 7.66                                                 & 26.57  & 2.36                                             & 0.55   & 4.49                                                 & 10.55  & 6.38                                             & 1.02   \\
Fixed\_3                 & 0.25                                                 & 1.41   & 21.58                                            & 0.31   & 3.84                                                 & 13.36  & 3.89                                             & 0.36   & 1.62                                                 & 4.60   & 10.53                                            & 0.48   \\ 
\bottomrule[1pt]
\end{tabular}
}
\end{table*}

\begin{figure*}[htp]
 \setlength{\belowcaptionskip}{-0.01cm}
\begin{minipage}{0.17\textwidth}
    \includegraphics[width=\textwidth]{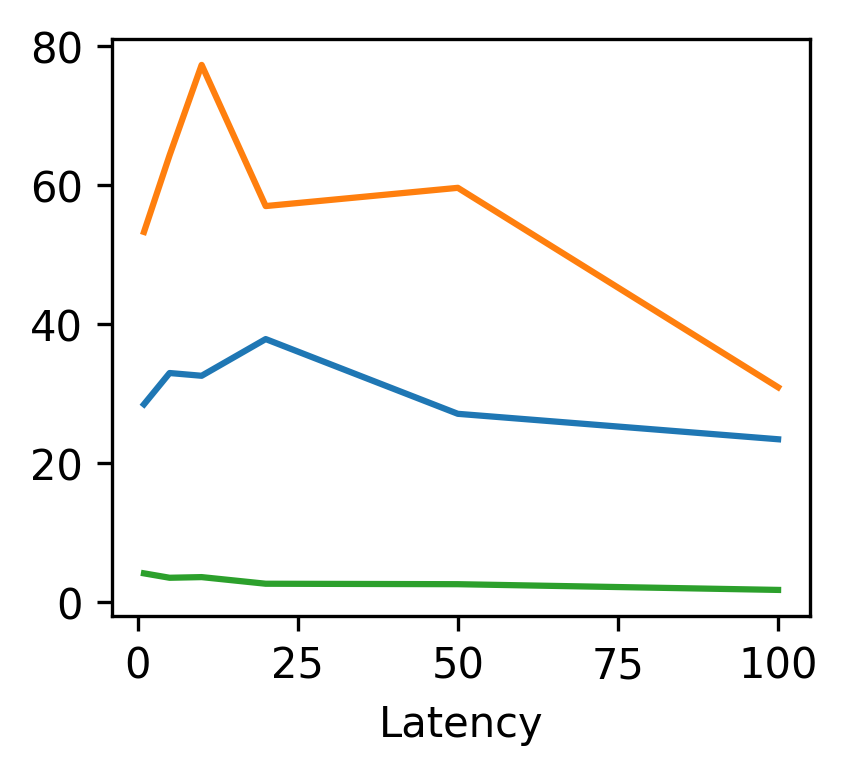}
    \centerline{(a) C-PPO}
\end{minipage}
\hfill
\begin{minipage}{0.17\textwidth}
    \includegraphics[width=\textwidth]{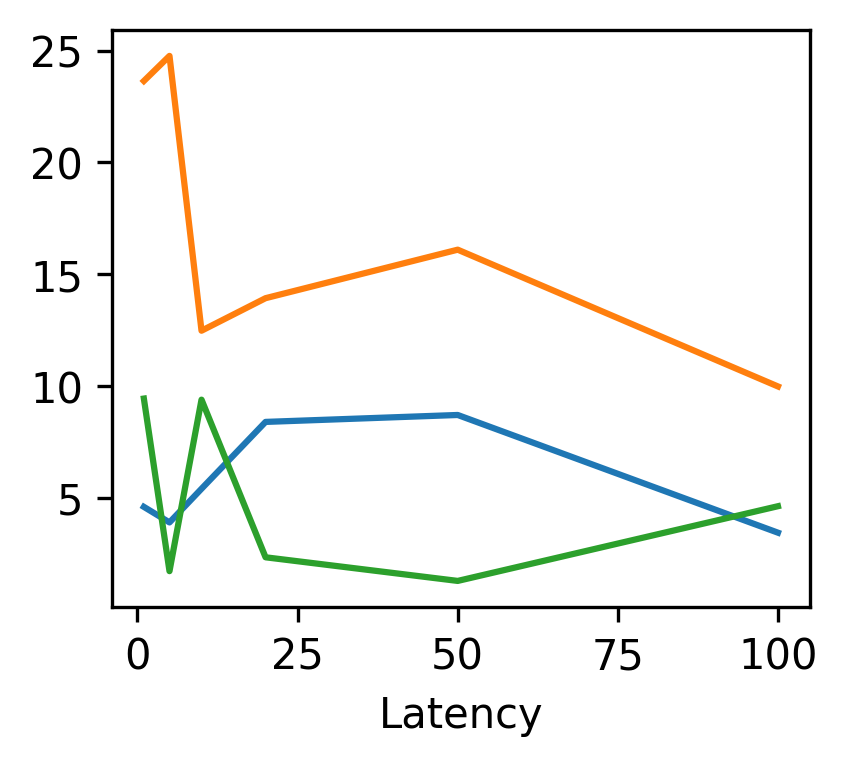}
    \centerline{(b) D-DQN}
\end{minipage}
\hfill
\begin{minipage}{0.17\textwidth}
    \includegraphics[width=\textwidth]{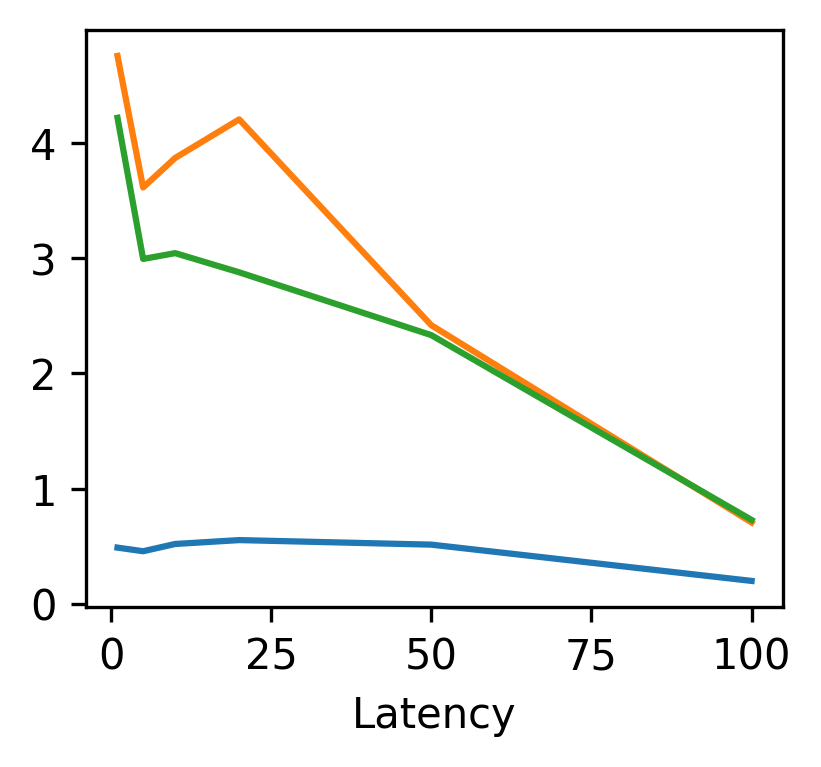}
    \centerline{(c) AS}
\end{minipage}
\hfill
\begin{minipage}{0.17\textwidth}
    \includegraphics[width=\textwidth]{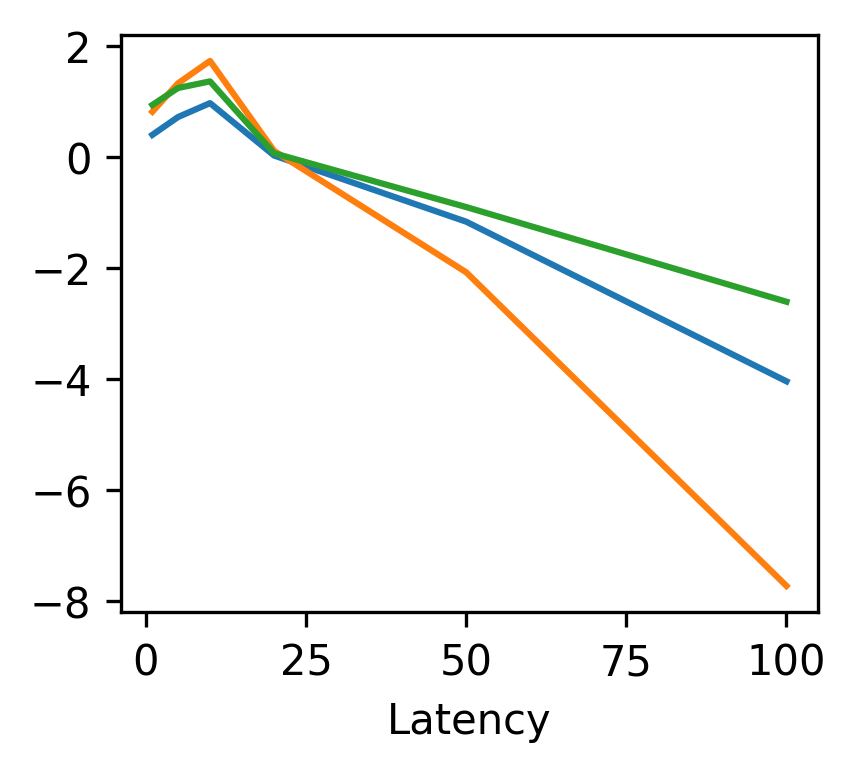}
    \centerline{(d) Random}
\end{minipage}
\hfill
\begin{minipage}{0.17\textwidth}
    \includegraphics[width=\textwidth]{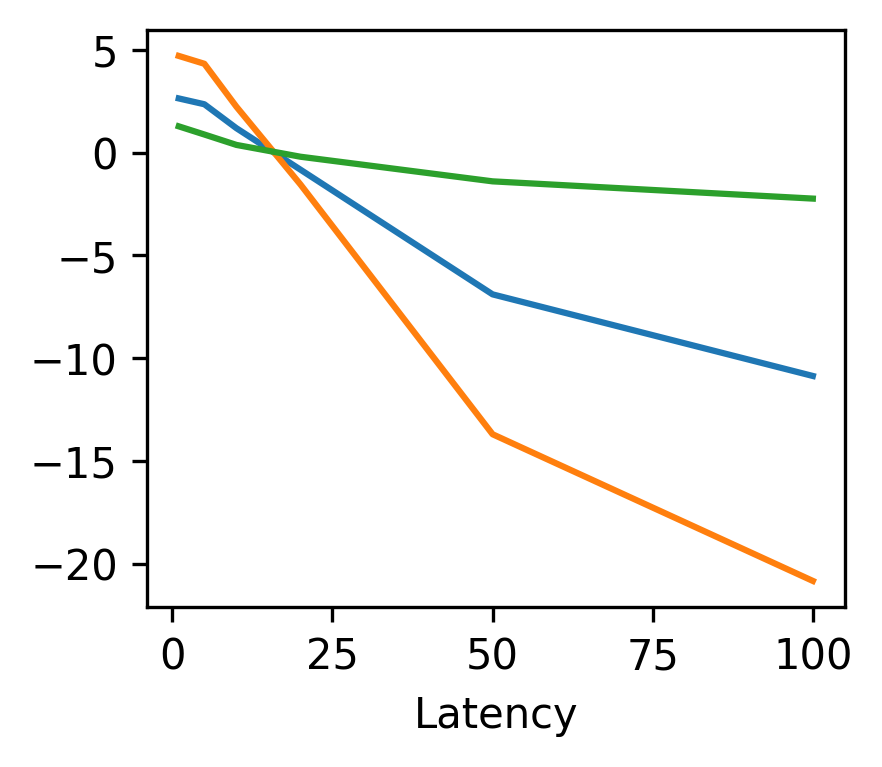}
    \centerline{(e) Fixed}
\end{minipage}
\hfill
\begin{minipage}{0.08\textwidth}
    \
    \includegraphics[width=\textwidth]{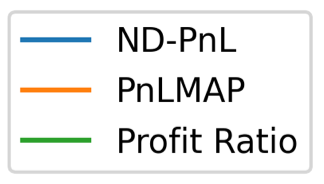}
\end{minipage}
\caption{Latency experiments.}
\label{fig:latency}
\end{figure*}

\subsubsection{Experimental Setup}
For the convenience of training and testing, we divide a trading day to some small episodes, each of which includes 2000 events, which is about 3-5 minutes. In each episode, the agent's value is initialized to 0, and at the end of each episode, the agent will close positions with market orders. The $PnL$ of this episode will be the final value of the agent. 
We split the train and test data as described in Section \ref{sec:exp_setup}. In the training phase, the agent can go through the training data once or more, while in the test phase, the agent can only go through it once.
We set hyperparameters $\omega=10$, $T=50$, $\eta=0.5$, $\zeta=0.01$.

Dueling DQN performed better in the discrete action space, so we use it to train the agent in the discrete action space. And PPO is used to train the agent in the continuous action space. The Discrete Dueling DQN is called D-DQN, The Continuous PPO is called C-PPO. The hyperparameters used in continuous action space are $max\_bias=0.05$, $max\_spread=0.1$.


\subsubsection{Compared Methods}
\begin{itemize}
    \item \textbf{Inventory-based RL} \cite{lim2018reinforcement} is an RL baseline, they used the agent's $inventory$ and $remaining\_time$ to represent market state.

    \item \textbf{LOB-based RL} \cite{zhong2020data} is another RL baseline which handcrafted market feature from limit order books. They used $bidSpeed$, $askSpeed$, $avgmidChangeFrac$, $invSign$, $cumPnL$ to represent the market.
    
    \item \textbf{Avellaneda-Stoikov Strategy} \cite{avellaneda2008high} is a classical model commonly used in the market making. It calculates the optimal quote price with a stochastic control model. 
    The reservation price and optimal spread are calculated by:
    \begin{equation}
        r(s,q,t) = s - q \gamma \sigma ^2(T-t) 
    \end{equation}
    \begin{equation}
        \delta ^a+\delta ^b=\gamma \sigma ^2(T-t) + \frac{2}{\gamma }\ln_{}{(1+\frac{\gamma }{\kappa } )}  
    \end{equation}
    where $s$ is the current mid-price, $q$ is the quantity of assets in inventory, $\sigma$ is a volatility parameter, $\gamma$ is a risk aversion parameter, $\kappa$ is a liquidity parameter, $T$ is the total time and $t$ is the current time.
    
    \item \textbf{Random Quoting Strategy} randomly quotes ask and bid orders in the five levels of the limit order book. 
    
    \item \textbf{Fixed Quoting Strategy} constantly quotes ask and bid orders in the fixed (1-3) level of the limit order book. 
\end{itemize} 

\subsubsection{Metrics}

\begin{itemize}
    \item \textbf{ND-PnL} \cite{spooner2018market} can rate how good our strategies are at capturing spreads. It is defined as PnL divided by the average spread in a period. It means how many spreads are captured by the agent on average.

    \item \textbf{PnLMAP} \cite{gavsperov2021market} is defined as PnL divided by the mean absolute position (MAP) in this period. It means the PnL in per unit of inventory and can measure the ability of the agent to profit against inventory risk.

    \item \textbf{Profit Ratio} (PR) is defined as the PnL divided by the total trading volume of the agent. 
    This metric measures the agent's profitability against transaction costs.

    \item \textbf{Sharpe ratio} 
    compares the return of an investment with its risk
    to evaluate the long-term profitability of a strategy. 
    
\end{itemize} 

\subsection{Overall Results}

As Table \ref{table:overall_results} shows, our RL agents achieve competitive performance especially in capturing spread and inventory controlling. C-PPO, in particular, beats all baselines, showing the advantage of the proposed continuous action space. 
For profit ratio, we notice that the smaller the quoting distance is, the lower the PR is, but the profitability increases (Fixed3 rendered the highest PR while the almost lowest ND-PnL and PnLMAP). The reason is that there are many useless transaction at the same price when quoting closely, so there is a trade-off.
For Sharpe ratio, we can see only our C-PPO and AS strategy still remain positive among all the targets, showing our stability of profits.
For capturing spreads, our C-PPO and D-DQN have a huge advantage. The best Fixed strategy also has a good performance (sometimes even defeats AS strategy), but it is not stable, whose optimal quoted distance differs from stocks (level 1 for \textit{Ping An Bank Co.,Ltd.} and level 2 for the other). And their inventory levels are higher than the RL methods. 
The biggest advantage of our agent is mainly on the PnLMAP metrics, it can obtain exceeding profit while maintain a rather low inventory level.
The AS strategy is also good at inventory controlling but at some expense of profitability. 

\subsection{Extended Experiments}
In this section, three extended experiments on latency, ablation and explanation are conducted in \textit{Ping An Bank Co.,Ltd.} to demonstrate the advantage of our method.

\subsubsection{Latency}

 \begin{table}[]
 \setlength{\abovecaptionskip}{-0.01cm}
 \setlength{\belowcaptionskip}{-0.01cm}
 \caption{The runtime of methods.}
\label{table:runtime}
\centering
\scalebox{0.9}{
\begin{tabular}{lccccccc}
\toprule[1pt]
\multirow{2}{*}{Method} & \multirow{2}{*}{Random} & \multirow{2}{*}{Fixed} & \multirow{2}{*}{AS} & \multicolumn{2}{c}{D-DQN}  & \multicolumn{2}{c}{C-PPO}           \\
                        &                         &                      &                     & Infer & Train                    & \multicolumn{1}{l}{Infer} & Train \\ \hline
Runtime (ms/ts)         & 10.0                    & 10.9                 & 19.7                & 46.7  & \multicolumn{1}{c}{77.5} & 47.5                      & 75.7  \\
\bottomrule[1pt] 
\end{tabular}
}
\end{table}

In HFT domain, latency, which is usually caused by messaging and calculation, is an important factor affecting the profitability of a strategy. This is because these quotes may not be based on the latest price information. 
Therefore, we are concerned about whether the agent can maintain its performance against latencies.
We investigated the robustness of our approach and baselines to latency, as shown in Figure \ref{fig:latency}. 

The AS, Random, and Fixed strategies suffer from latencies, which is because these models' parameters are fixed and therefore cannot accommodate latencies. And our RL agents, C-PPO and D-DQN, perform better than them, because the agent may self-adapt to latencies in the environment. 
Since our pre-training horizon is equal to 10, it performs better when the latency is around 10.

And we record the running time of different methods, as shown in Table \ref{table:runtime}. Baselines have a huge advantage in running time, but considering the average interval between events is 60-150 ms, the computation time of our agents is acceptable.


\subsubsection{Ablation}

\begin{table}[]
\setlength{\abovecaptionskip}{-0.01cm}
 \setlength{\belowcaptionskip}{-0.01cm}
\caption{Ablation experiments.}
\label{table:ablation}
\centering
\scalebox{0.95}{
\begin{tabular}{lcccc}
\toprule[1pt]
                      & \begin{tabular}[c]{@{}c@{}}ND-PnL\\ $(\times 10^{5})$\end{tabular} & PnLMAP & \begin{tabular}[c]{@{}c@{}}Profit Ratio\\ $(\times 10^{-4})$\end{tabular} & Sharpe \\ 
\toprule[1pt]
\textbf{C-PPO}              & 9.34                                            & 117.18       & 5.01                                                       & 12.34   \\ \hline
\qquad w/o LOB state               & 0.15                                             & 17.13        & 10.66                                                       & 1.20   \\
\qquad w/o Attn-LOB                & 0.58                                            & 22.18       & 11.15                                                       & 1.43   \\
\qquad w/o Dynamic state           & 9.12                                           & 112.52       & 5.05                                                       & 13.32   \\ 
\bottomrule[1pt]
\textbf{D-DQN}              & 6.98                                             & 8.65       & 3.54                                                       & 1.25   \\ \hline
\qquad w/o LOB state               & 4.52                                             & 6.60       &1.70                                                       & 2.71   \\
\qquad w/o Attn-LOB                & 6.96                                             & 7.83        & 3.46                                                       & 1.31   \\
\qquad w/o Dynamic state           & 6.38                                             & 7.90       & 3.95                                                       & 1.11   \\  
\bottomrule[1pt]
\end{tabular}
}
\end{table}


In order to investigate the effects of each part in our model, we perform ablation experiments, and the results are shown in Table \ref{table:ablation}. 

The rich microstructure information contained in limit order books are key features to a market making agent. If we do not input the LOB state, the performance decreased severely.
But it can only be extracted by a specially designed module, if we replace Attn-LOB with a simple MLP, the performance only increase slightly.
The dynamic state, which represents the market dynamics, also enhanced the performance, but not so much.


\subsubsection{Explanation}

\begin{figure}[htp]
 \setlength{\belowcaptionskip}{-0.1cm}
\begin{minipage}{0.23\textwidth}
    \includegraphics[width=\textwidth]{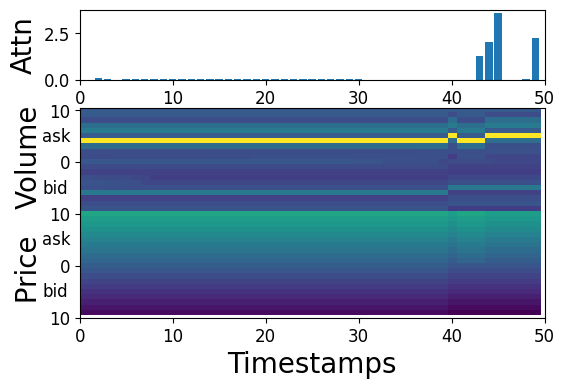}
\end{minipage}
\hfill
\begin{minipage}{0.23\textwidth}
    \includegraphics[width=\textwidth]{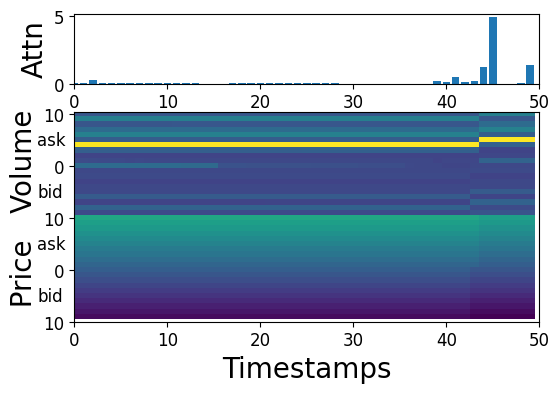}
\end{minipage}
\centerline{(a) Stable markets.}
\vfill
\begin{minipage}{0.23\textwidth}
    \includegraphics[width=\textwidth]{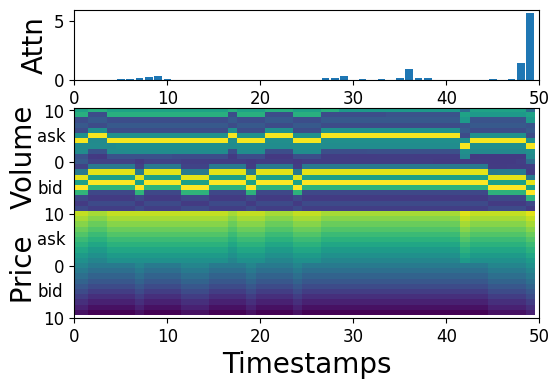}
\end{minipage}
\hfill
\begin{minipage}{0.23\textwidth}
    \includegraphics[width=\textwidth]{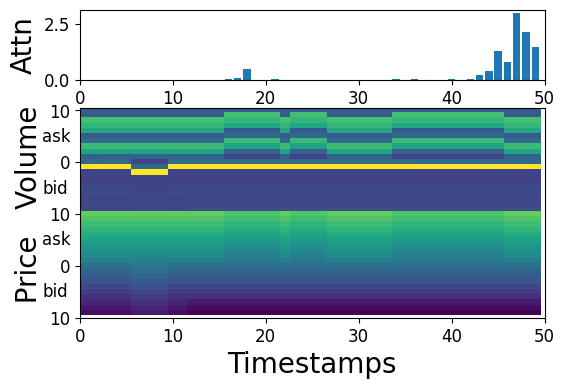}
\end{minipage}
\centerline{(b) Rapidly changing markets.}
\caption{Attention visualization.}
\label{fig:attn}
\end{figure}

By visualizing the weight of self-attention layer in Figure \ref{fig:attn}, we find that the agent paid the most attention to the most recent events. In a stable market like (a), the agent will pay attention to the latest change, while in a rapidly changing market like (b), it will also look at the earlier market changes.

\begin{figure}[h]
 \setlength{\belowcaptionskip}{-0.1cm}
  \centering
  \includegraphics[width=5cm]{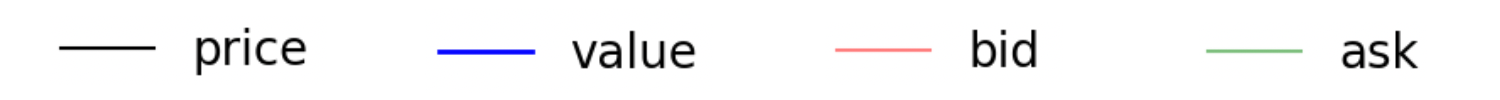}
  \includegraphics[width=8cm]{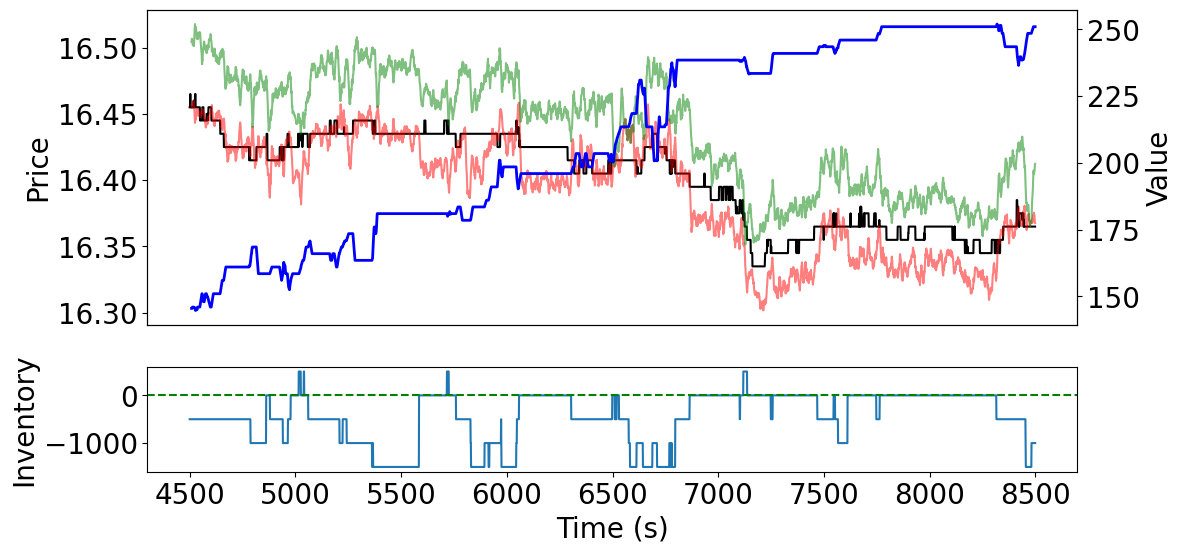}
  \caption{An example of agent decision making.}
  \label{fig:trade}
\end{figure}

In Figure \ref{fig:trade}, we plot historical decisions of the C-PPO agent to show that it did learn some trading guidelines. In the first half of the picture, where the market fluctuated around a certain price, the agent made profits by opening and closing positions. When the agent holds a negative position, it puts the bid quote very near the mid-price in an effort to revert to a neutral position. 
However, in the second half of the picture, the price fell sharply and a trending market emerged. The quoted prices are far away from the mid-price, so as to keep a low inventory position to avoid risks.

%% file: sections/conclusion.tex
\section{Conclusion}

In this paper, we propose a novel deep RL agent for market making, which shows the competitive ability for capturing spreads and controlling inventory, and the robustness to latencies. 
To achieve this, we leverage a pre-trained neural network Attn-LOB to extract features from limit order books. Then we propose a novel continuous action space and a hybrid reward function, and we show their effectiveness with ablation experiments. 
We explain our results by visualizing attention weights. The agent paid the most attention to the nearest events while still looking at very early changes. And we find that the agent can learn some market-making skills like humans.
Empirically, the proposed agent achieved satisfactory performance, outperforming the traditional quantitative model and RL baselines on the simulated environment of three stocks.